\documentclass[epj]{webofc}
\usepackage[utf8]{inputenc}
\usepackage[varg]{txfonts}   
\usepackage{booktabs}
\usepackage{xcolor}
\usepackage{verbatim}

\newcommand{\be}{\begin{equation}}
\newcommand{\ee}{\end{equation}}
\newcommand{\bea}{\begin{eqnarray}}
\newcommand{\eea}{\end{eqnarray}}

\newcommand{\bie}{\begin{small} \begin{itemize}}

\newcommand{\eie}{\end{itemize} \end{small}}

\definecolor{darkred}{rgb}{0.4,0.0,0.0}
\definecolor{darkgreen}{rgb}{0.0,0.4,0.0}
\definecolor{darkblue}{rgb}{0.0,0.0,0.4}
\usepackage[bookmarks,linktocpage,colorlinks,
    linkcolor = darkred,
    urlcolor  = darkblue,
    citecolor = darkgreen]{hyperref}
\usepackage{braket}
\usepackage[english]{babel}  
\usepackage[utf8]{inputenc}   
\usepackage[sans]{dsfont}
\usepackage{eurosym}
\usepackage{array}
\usepackage{tabularx}
\usepackage{booktabs}
\usepackage{bm}
\usepackage{amsmath}
\usepackage{amssymb}
\usepackage{amsmath, amsfonts, epsfig, xspace}
\usepackage{listings}
\usepackage{algorithmic}
\usepackage{verbatim}
\usepackage{grffile}
\usepackage{indentfirst}

\usepackage{subfig}
\usepackage{float}
\usepackage{url}

\usepackage{adjustbox}
\usepackage{booktabs}
\usepackage{multirow}

\usepackage{cleveref}
\crefname{section}{Section}{Sections}
\Crefname{section}{Section}{Sections}
\crefname{equation}{Eq.}{Eqs.}
\Crefname{equation}{Eq.}{Eqs.}
\crefname{figure}{Figure}{Figs.}
\Crefname{figure}{Figure}{Figs.}
\crefname{table}{Table}{Tables}
\Crefname{table}{Table}{Tables}

\newcommand{\Tr}{\mbox{\rm Tr}}

\usepackage{ragged2e}

\newcommand{\LagFigSize}{0.85\columnwidth}

\usepackage{ragged2e}

\graphicspath{{fig/}}

\wocname{EPJ Web of Conferences}
\woctitle{Lattice2017}
\begin{document}
%
\selectlanguage{english}
\title{%
Colour fields of the quark-antiquark excited flux tube 
}
\author{%
\firstname{Pedro} \lastname{Bicudo}\inst{1}\fnsep\thanks{Speaker, \email{bicudo@tecnico.ulisboa.pt} } 
\and
\firstname{Marco} \lastname{Cardoso}\inst{1} 
\and
\firstname{Nuno}  \lastname{Cardoso}\inst{1}
}
\institute{%
Portuguese Lattice QCD Collaboration 
and Departamento de F\'{\i}sica, Instituto Superior T\'ecnico,
Universidade de Lisboa, Avenida Rovisco Pais 1, 1049-001 Lisboa, Portugal
}
\abstract{%
We present colour field density profiles for some of the first SU(3) gluonic excitations of the flux
tube in the presence of a static quark-antiquark pair. The results are obtained from a large set of
gluonic operators. 
 }
\maketitle

\section{Introduction}\label{intro}

Understanding the confinement of colour remains a main theoretical problems of modern physics. Its solution could also open the door to other unsolved theoretical problems. One of the evidences of confinement, where we may search for clues to solve it, is in the QCD flux tubes \cite{Wilson:1974sk}. 

However the details of confinement are masked by the dominant string-like behaviour of the flux tubes, with a single scale $\sigma$, independent of the details of confinement. 
The main analytical model utilized in the literature to explain the behaviour of the QCD flux tubes is the Nambu-Goto string model. It assumes infinitely thin strings, with transverse quantum fluctuations only. The quantum fluctuations predict not only a zero mode width of the groundstate flux tube, increasing with distance, but also an infinite tower of quantum excitations \cite{Juge:2002br}. Both effects have been observed by lattice QCD computations, indeed confirming the string dominance of the QCD flux tube.

Nevertheless recently, our lattice QCD collaboration  PtQCD \cite{PtQCD} studied the zero temperature groundstate flux tube of pure gauge QCD, and found evidence for a penetration length $\lambda$ \cite{Cardoso:2013lla}, as a second scale other than the string tension $\sigma$, contributing to the colour fields density profile of the flux tube.

Here we study quantitatively the excitations of the QCD flux tube. Our goal is to eventually compare the flux tubes with the densities expected in the Nambu-Goto string model. We think it is important to search for deviations from the model, which only considers transverse quantum excitations. Moreover, the excitations of the flux tube, in mesonic systems, may lead to hybrid mesons, which also remain to be understood.

We apply lattice QCD computations, in a first preliminary study, to explore the colour field density profiles of the quantum excitations of the pure gauge QCD flux tube produced by a a static quark and a static antiquark. We utilize the correlation matrix of Wilson loops with a large set of flux tube operators. To have enough computer power, we write our codes in CUDA and run them in computer servers with NVIDIA GPUs.

\section{Our 33 operator basis to produce the different excited quantum numbers \label{sec:operator}}

\begin{figure}[t!]
\begin{centering}
\includegraphics[width=0.7\columnwidth]{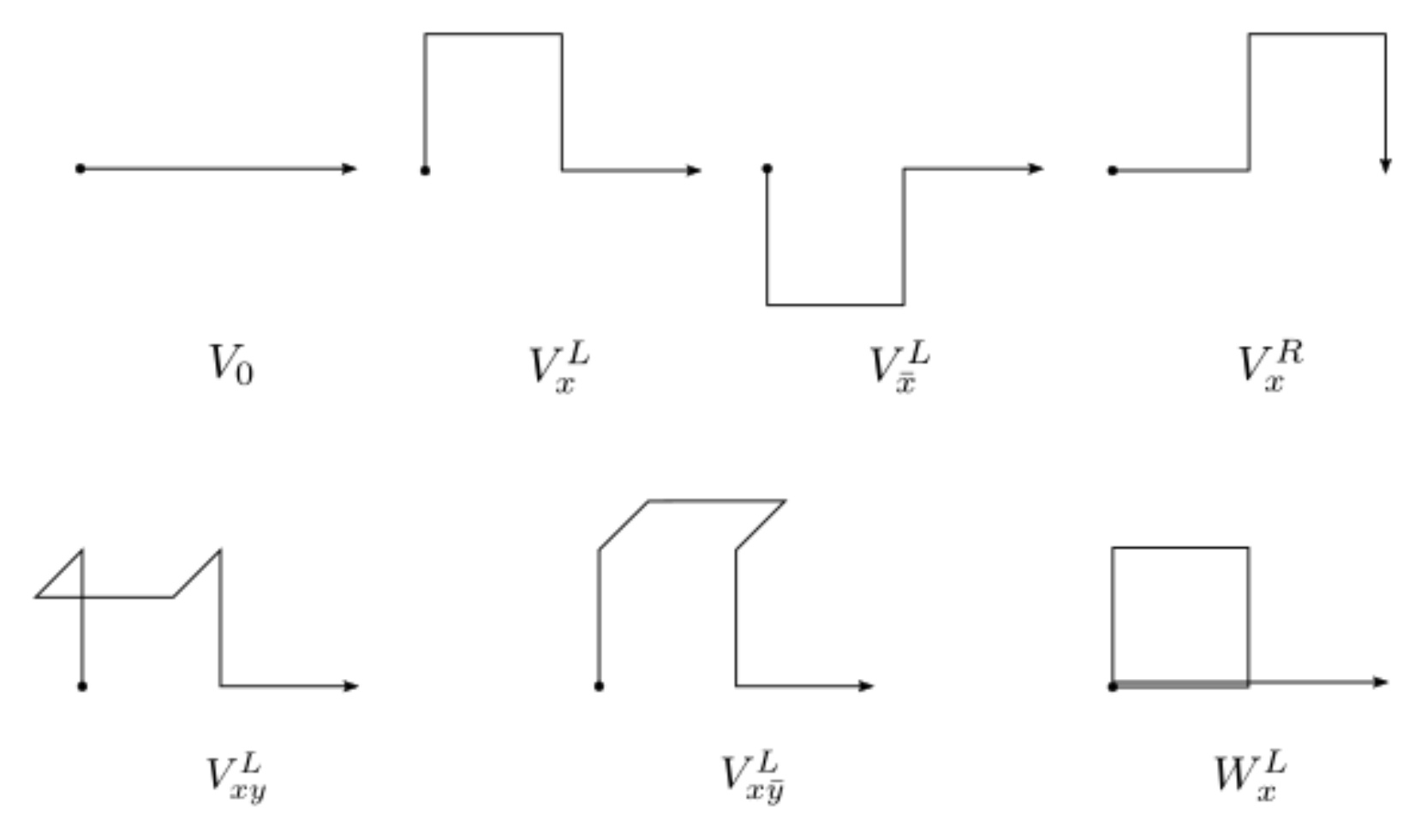}\caption{Examples of the paths from the quark to the antiquark used to construct the gauge field operators.
\label{fig:basisex}}
\par\end{centering}
\end{figure}

In the study of the flux tubes, we utilize a basis of spacial Wilson line operators, defined in Figure  \ref{fig:basisex}, sufficiently complete to include different types of flux tube excitations \cite{Lacock:1996vy}.
Our basis is composed by four kinds of operators:
\begin{itemize}
\item The direct operator $V_{0}$.
\item The eight open-staple operators $V_{x}^{L}$ , $V_{y}^{L}$, $V_{\bar{x}}^{L}$,
$V_{\bar{y}}^{L}$, $V_{x}^{R}$, $V_{y}^{R}$, $V_{\bar{x}}^{R}$
and $V_{\bar{y}}^{R}$.
\item The sixteen open-staple two-direction operators $V_{xy}^{L}$, $V_{x\bar{y}}^{L}$,
$V_{\bar{x}y}^{L}$, $V_{\bar{x}\bar{y}}^{L}$, $V_{yx}^{L}$, $V_{y\bar{x}}^{L}$,
$V_{\bar{y}x}^{L}$, $V_{\bar{y}\bar{x}}^{L}$, $V_{xy}^{R}$, $V_{x\bar{y}}^{R}$,
$V_{\bar{x}y}^{R}$, $V_{\bar{x}\bar{y}}^{R}$, $V_{yx}^{R}$, $V_{y\bar{x}}^{R}$,
$V_{\bar{y}x}^{R}$ and $V_{\bar{y}\bar{x}}^{R}$.
\item The eight closed-staple operators similar to the open-staple ones
$W_{x}^{L}$, $W_{y}^{L}$, $W_{\bar{x}}^{L}$,
$W_{\bar{y}}^{L}$, $W_{x}^{R}$, $W_{y}^{R}$, $W_{\bar{x}}^{R}$ and $W_{\bar{y}}^{R}$.
\end{itemize}
The bar means that there is displacement in the negative axis direction.
The $L$ and $R$ labels indicate whether the staple is on the left
or on the right.

%

This amounts in 33 different operators. Although it would be interesting to use a more complete basis with more operators, but since the computation of the flux tube profiles is extremely demanding, we limit our basis to the present 33 operators. 

To further limit the size of the correlation matrix, we first block diagonalize it.
With linear combinations of our 33 operators, we construct operators with a definite symmetry. 
We utilize the same quantum number notation of Ref. \cite{Juge:2002br}, 
see also the recent Ref. \cite{Reisinger:2017btr}. 
There are three symmetry groups for a flux tube between two static sources, and they determine three quantum numbers. 
\begin{itemize}
\item
The two-dimensional rotation along the charge axis has angular number $\Lambda = {\bf J}_g\!\cdot\hat{\bf R}$ projected in the unit vector
$\hat{\bf R}$. The capital Greek
letters $\Sigma, \Pi, \Delta, \Phi, \dots$ indicate as usually states
with $\Lambda=0,1,2,3,\dots$, respectively.  
\item 
The permutation of the quark and the antiquark static charges is equivalent to a combined operations of
charge conjugation and spatial inversion about the origin, which we set at the midpoint between the
quark and the antiquark is also a symmetry. Its eigenvalue is denoted by
$\eta_{CP}$.  States with $\eta_{CP}=1 (-1)$ are denoted
by the subscripts $g$ ($u$).  
\item 
Due to the planar, and not three-dimensional, angular momentum there is an additional label for the
$\Sigma$ states. $\Sigma$ states which
are even (odd) under a reflection in a plane containing the molecular
axis are denoted by a superscript $+$ $(-)$. This is different from the phase corresponding to a two-dimensional p-wave. 
\end{itemize}

  Hence, the low-lying
levels are labeled $\Sigma_g^+$, $\Sigma_g^-$, $\Sigma_u^+$, $\Sigma_u^-$,
$\Pi_g$, $\Pi_u$, $\Delta_g$, $\Delta_u$, and so on.  For convenience,
we use $\Gamma$ to denote these labels in general.
As a result of the different symmetries and respective quantum numbers, we rearrange our initial 33 operators into the following operators.
\begin{itemize}
\item $\bm{\Sigma_{g}^{+}}$
\end{itemize}

For this quantum number, we have four operators:
\begin{eqnarray*}
\mathcal{A}_{0,1} & = & V_{0}\\
\mathcal{A}_{0,2} & = & \frac{1}{2\sqrt{2}}\big(V_{x}^{L}+V_{y}^{L}+V_{\bar{x}}^{L}+V_{\bar{y}}^{L}+V_{x}^{R}+V_{y}^{R}+V_{\bar{x}}^{R}+V_{\bar{y}}^{R}\big)\\
\mathcal{A}_{0,3} & = & \frac{1}{4}\big(V_{xy}^{L}+V_{x\bar{y}}^{L}+V_{\bar{x}y}^{L}+V_{\bar{x}\bar{y}}^{L}+V_{yx}^{L}+V_{y\bar{x}}^{L}+V_{\bar{y}x}^{L}+V_{\bar{y}\bar{x}}^{L}\\
 &  & \ +V_{xy}^{R}+V_{x\bar{y}}^{R}+V_{\bar{x}y}^{R}+V_{\bar{x}\bar{y}}^{R}+V_{yx}^{R}+V_{y\bar{x}}^{R}+V_{\bar{y}x}^{R}+V_{\bar{y}\bar{x}}^{R}\big)\\
\mathcal{A}_{0,4} & = & \frac{1}{2\sqrt{2}}\big(W_{x}^{L}+W_{y}^{L}+W_{\bar{x}}^{L}+W_{\bar{y}}^{L}+W_{x}^{R}+W_{y}^{R}+W_{\bar{x}}^{R}+W_{\bar{y}}^{R}\big)
\end{eqnarray*}

\begin{itemize}
\item $\bm{\Sigma_{u}^{+}}$
\end{itemize}

For this quantum number, we have three operators:
\begin{eqnarray*}
\mathcal{A}_{2,1} & = & \frac{1}{2\sqrt{2}}\big(V_{x}^{L}+V_{y}^{L}+V_{\bar{x}}^{L}+V_{\bar{y}}^{L}-(V_{x}^{R}+V_{y}^{R}+V_{\bar{x}}^{R}+V_{\bar{y}}^{R})\big)\\
\mathcal{A}_{2,2} & = & \frac{1}{4}\big(V_{xy}^{L}+V_{x\bar{y}}^{L}+V_{\bar{x}y}^{L}+V_{\bar{x}\bar{y}}^{L}+V_{yx}^{L}+V_{y\bar{x}}^{L}+V_{\bar{y}x}^{L}+V_{\bar{y}\bar{x}}^{L}\\
 &  & \ -(V_{xy}^{R}+V_{x\bar{y}}^{R}+V_{\bar{x}y}^{R}+V_{\bar{x}\bar{y}}^{R}+V_{yx}^{R}+V_{y\bar{x}}^{R}+V_{\bar{y}x}^{R}+V_{\bar{y}\bar{x}}^{R})\big)\\
\mathcal{A}_{2,3} & = & \frac{1}{2\sqrt{2}}\big(W_{x}^{L}+W_{y}^{L}+W_{\bar{x}}^{L}+W_{\bar{y}}^{L}-(W_{x}^{R}+W_{y}^{R}+W_{\bar{x}}^{R}+W_{\bar{y}}^{R})\big)
\end{eqnarray*}

\begin{itemize}
\item $\bm{\Pi_{u}}$
\end{itemize}

We have four operators for these quantum numbers:
\begin{eqnarray*}
\mathcal{A}_{4,1} & = & \frac{1}{2\sqrt{2}}\big(V_{x}^{L}+iV_{y}^{L}-V_{\bar{x}}^{L}-iV_{\bar{y}}^{L}+V_{x}^{R}+iV_{y}^{R}-V_{\bar{x}}^{R}-iV_{\bar{y}}^{R}\big)
\\
\mathcal{A}_{4,2} & = & \frac{1}{4}\big(V_{xy}^{L}+V_{x\bar{y}}^{L}-V_{\bar{x}y}^{L}-V_{\bar{x}\bar{y}}^{L}+iV_{yx}^{L}+iV_{y\bar{x}}^{L}-iV_{\bar{y}x}^{L}-iV_{\bar{y}\bar{x}}^{L} 
\\ && 
+V_{xy}^{R}+V_{x\bar{y}}^{R}-V_{\bar{x}y}^{R}-V_{\bar{x}\bar{y}}^{R}+iV_{yx}^{R}+iV_{y\bar{x}}^{R}-iV_{\bar{y}x}^{R}-iV_{\bar{y}\bar{x}}^{R}\big) 
\\
\mathcal{A}_{4,3} & = & \frac{1}{4}\big(V_{xy}^{L}-V_{x\bar{y}}^{L}+V_{\bar{x}y}^{L}-V_{\bar{x}\bar{y}}^{L}-iV_{yx}^{L}+iV_{y\bar{x}}^{L}-iV_{\bar{y}x}^{L}+iV_{\bar{y}\bar{x}}^{L} 
\\ &&
+V_{xy}^{R}-V_{x\bar{y}}^{R}+V_{\bar{x}y}^{R}-V_{\bar{x}\bar{y}}^{R}-iV_{yx}^{R}+iV_{y\bar{x}}^{R}-iV_{\bar{y}x}^{R}+iV_{\bar{y}\bar{x}}^{R}\big)\\
\mathcal{A}_{4,4} & = & \frac{1}{2\sqrt{2}}\big(W_{x}^{L}+iW_{y}^{L}-W_{\bar{x}}^{L}-iW_{\bar{y}}^{L}+W_{x}^{R}+iW_{y}^{R}-W_{\bar{x}}^{R}-iW_{\bar{y}}^{R}\big)
\end{eqnarray*}

In what concerns the operators with quantum numbers  $\bm{\Sigma_{g}^{-}}$, we do not list them here since they have a high energy and we did not get a clear enough signal for the respective flux tube.

\section{Computation of the excited state spectra \label{sec:spectra}}

We start by utilizing the correlation matrix $\langle {\mathcal W}_{kl}(t) \rangle$ to compute the energy levels of the excited states, as done previously in the literature. 
Now the subindices $k$ and $l$ stand for the spacial operators in the operator basis defined in Section  \ref{sec:operator}, now denoted $O_k$.
The spacial operators are connected by temporal Wilson lines $L$,
\begin{equation}
 {\mathcal W}_{kl}(t)=
 O_k(-\mathbf R/2, \mathbf R/2,-t/2) \, 
 L(\mathbf R /2, -t/2, t/2) \, 
 O_l^\dagger(\mathbf R/2, - \mathbf R /2, t/2) \,   
 L^\dagger(\mathbf R /2, t/2, -t/2) \ .
\end{equation}
The statistical average $\langle \cdots \rangle$ is performed over our ensemble of gauge link configurations.

Notice each matrix element corresponds to an evolution operator in Euclidian space, where all energy levels $E_i$ contribute, with coefficients depending on how close the operator is to the actual physical states, with the  Euclidian damping factor $\exp( -E_i \, t)$.

The first step to compute the energy levels, is to diagonalize the correlation matrix 
$\langle {\mathcal W}_{kl}(t) \rangle$, for each time extent $t$ of the Wilson loop, and get a set of time dependent eigenvalues $\lambda_i(t)$.
With the time dependence, we study the effective mass plot
\begin{equation}
E_i \simeq - \log { \lambda_i(t+1) \over \lambda_i(t)} \ ,
\end{equation}
and search for clear plateaus consistent with a constant energy $E_i$ in intervals $ t \in [t_\text{ini}, t_\text{fin}]$ between the initial and final time of the plateau.
The different energies $E_i$ levels, should correspond to the groundstate and excited states of the flux tube. If our operator basis is good enough, then $E_0$ is extremely close to the groundstate energy, $E_1$ is very close to the the first excited state, etc.

Moreover, with the diagonalization 
we also obtain the eigenvector operators corresponding to the groundstate, first excitation, etc. 
We get a linear combination of our initial operators,
\begin{eqnarray}
\label{eq:eigen}
\widetilde O_0=a_{01}\, O_1+ a_{02} \, O_2 + \cdots
\\
\nonumber
\widetilde O_1=a_{11} \, O_1+ a_{12} \, O_2 + \cdots
\\
\nonumber
\cdots
\end{eqnarray}
Notice this result must be interpreted with a grain of salt. 
The eigenvector operators $\widetilde O_i$ do not exactly correspond to the respective state as in quantum mechanics,
but they get the clearest possible signal to noise ratio, among our operator basis.

The eigenvector operators $\widetilde O_i$ and the respective correlation matrix can be used in the same time interval $ t \in [t_\text{ini}, t_\text{fin}]$ ideal for the effective mass plateaus of the energy spectrum.

\section{Computation of the chromofields in the flux tube}

As in Ref. \cite{Cardoso:2013lla}, the central observables that govern the event in the flux tube can be extracted from the correlation of a plaquette, $\square_{\mu\nu}$, with the quark-antiquark Wilson loops $\cal W$,
\begin{equation}
f_{\mu\nu}(R,x) = \frac{\beta}{a^4} \left[\frac{\Braket{\mathcal{W}(\mathbf R,t)\,\square_{\mu\nu}(\mathbf r)}}{\Braket{\mathcal{W}(\mathbf R,t)}}-\Braket{\square_{\mu\nu}(\mathbf r)}\right]
\label{eq:field}
\end{equation}
where $\mathbf r=(x,y,z)$ denotes the spacial distance of the plaquette from the centre of line connecting quark sources, $R$ is the quark-antiquark separation and $t$ is the time extent of the Wilson loop.

Therefore, using the plaquette orientation $(\mu,\nu)=(2,3), (1,3), (1,2),$ $(1,4),(2,4),(3,4)$, we can relate the six components in \cref{eq:field} to the components of the chromoelectric and chromomagnetic fields,
\begin{equation}
f_{\mu\nu}\rightarrow\left(\Braket{B_x^2},\Braket{B_y^2},\Braket{B_z^2},\Braket{E_x^2},\Braket{E_y^2},\Braket{E_z^2}\right) \ .
\end{equation}
Notice these are the Euclidan space components. In Minkowski space we must add a $-$ phase to the magnetic field density, $B_i^2 \to - B_i^2$.  With the field densities it is then trivial to compute  the total action (Lagrangian) density, $\Braket{\mathcal{L}}=\frac12\left(\Braket{E^2}+\Braket{B^2}\right)$.

In order to improve the signal over noise ratio, we use multihit technique in the temporal Wilson lines and the APE smearing spatial Wilson lines.
The multihit technique,  \cite{Brower:1981vt, Parisi:1983hm}, replaces each temporal link by its thermal average,
\begin{equation}
	U_4\rightarrow \bar{U}_4=\frac{\int dU_4 U_4 \,e^{\beta\Tr \left[U_4 F^\dagger\right]}}{\int dU_4 \,e^{\beta\Tr\left[ U_4 F^\dagger\right]}}
\end{equation}

Now, to extend Eq. (\ref{eq:field}) for the study of excited flux tubes, we simply have to replace the Wilson loop $\cal W$ by $ \widetilde {\cal{W}}_i$, where the spacial links are given by the eigenvector operators $\widetilde O_i$ of Eq. (\ref{eq:eigen}).

The eigenvector operators $\widetilde O_i$ and the respective Wilson loop $ \widetilde {\cal{W}}_i$ can be used in the same time interval $ t \in [t_\text{ini}, t_\text{fin}]$ ideal for the effective mass plateaus of the energy spectrum.

%
\begin{figure}[t!]
\begin{centering}
\includegraphics[width=0.7\columnwidth]{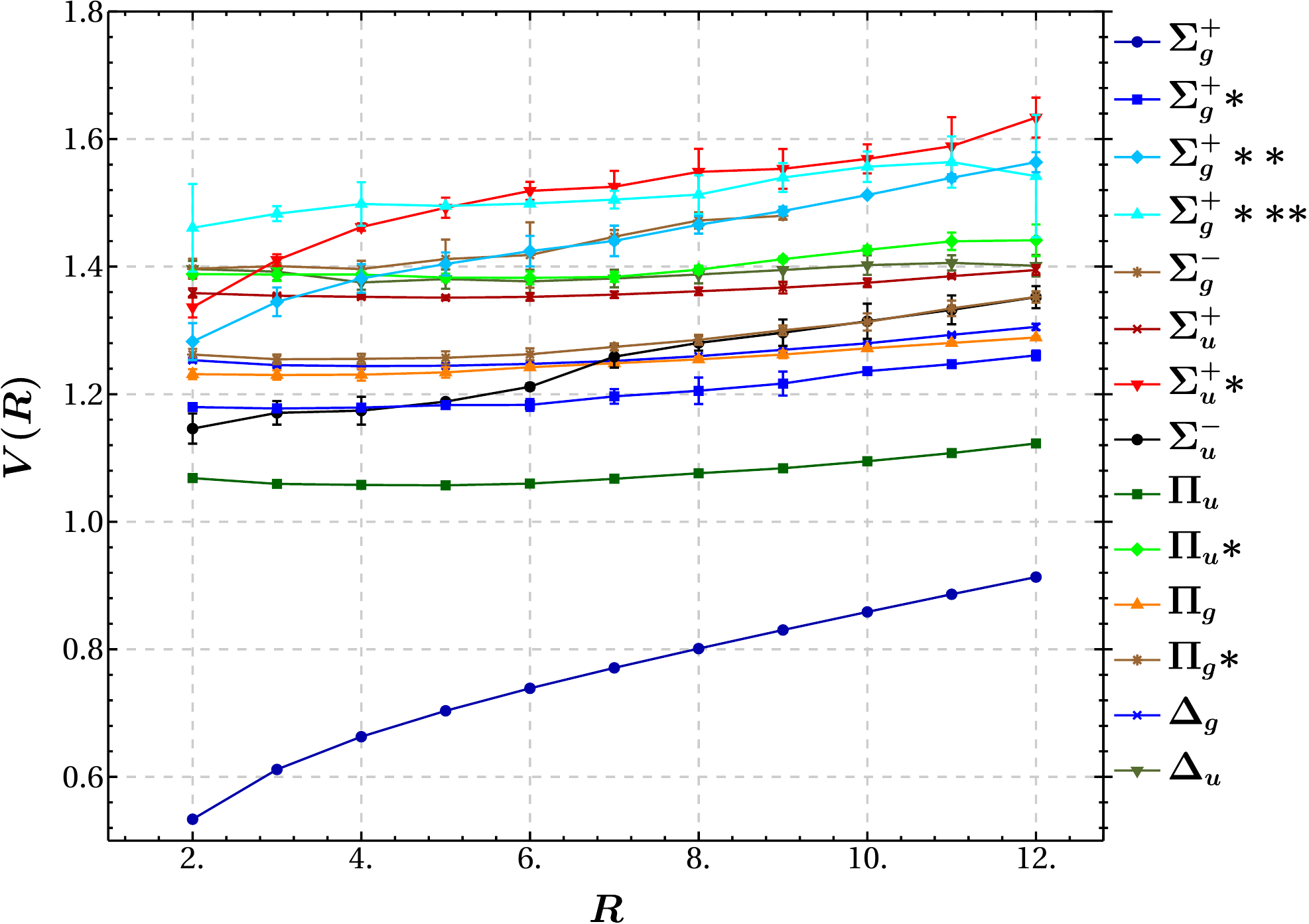}
\par\end{centering}
\caption{Flux tube spectra $V(R)$  as a function of the charge distance $R$. Distance and the energy are shown in lattice spacing units $a$.
\label{fig:pot}}
\end{figure}

\section{Configuration ensemble and code efficiency}

In this section, we present the results using 1199 configurations for a fixed lattice volume of $24^3\times 48$ and $\beta=6.2$. 
We present our results in lattice spacing units of a, with $a = 0.07261(85)\, \text{fm}$ or $a^{-1} = 2718(32)\, \text{MeV}$.
The the quark and antiquark are located at $(0, 0, -R/2)$ and $(0, 0, R/2)$ for $R$ between 6 and 10 in lattice spacing units.

All our computations are performed in NVIDIA GPUs using our CUDA codes. The computation of the chromofields are very computer intensive and due to the GPU limited memory this requires an intensive use of atomic memory operations. For example, to calculate the fields for the $\Sigma_u^+$ in the GeForce GTX TITAN Black, it takes approximately 83 minutes.

\section{Results}

In Figure \ref{fig:pot} we show our results for the flux tube spectra, as a function of the charge distance $R$. Distance and the energy are shown in lattice unit $a$ units. The ground-state $\Sigma^+_g$ is the familiar static-quark potential \cite{Cardoso:2013lla}. The lowest-lying excitation is the $\Pi_u$.

\begin{figure}[t!]
\begin{centering}
\includegraphics[width=\LagFigSize]{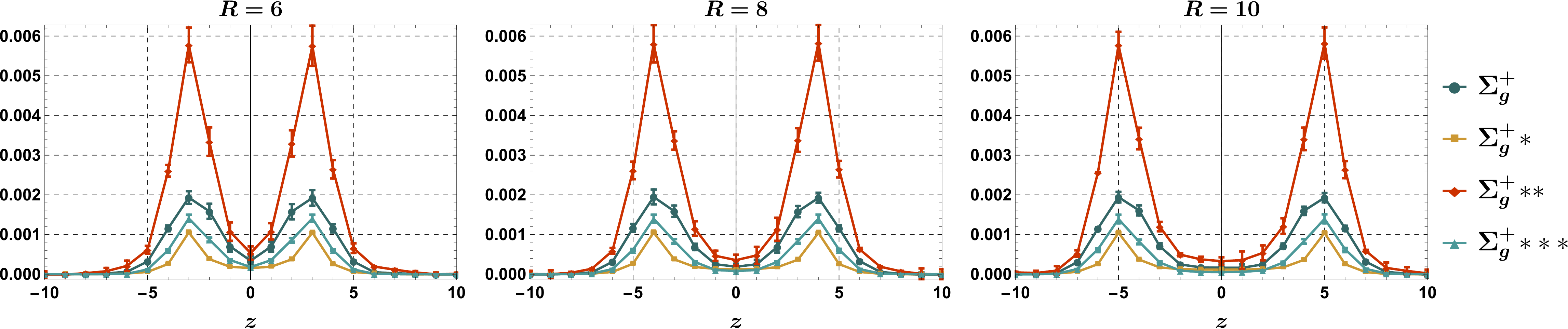}

\includegraphics[width=\LagFigSize]{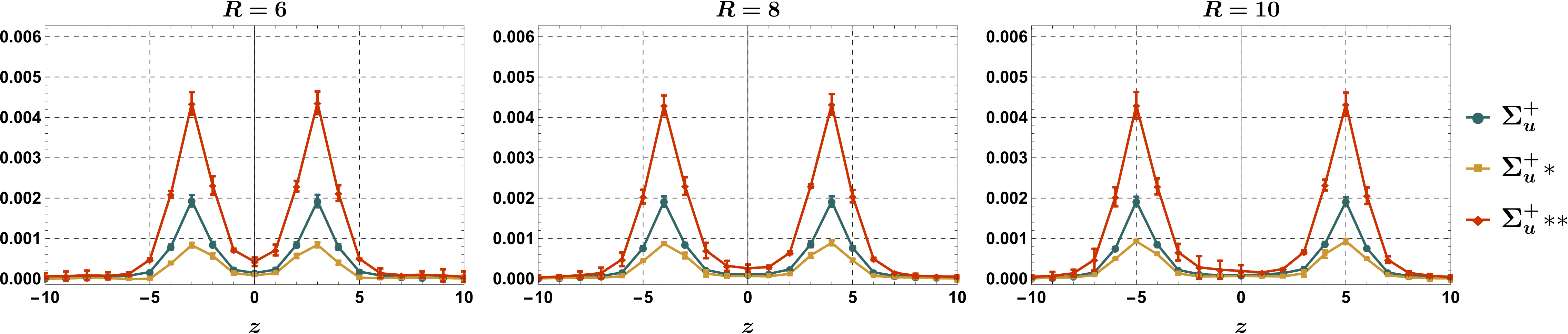}

\includegraphics[width=\LagFigSize]{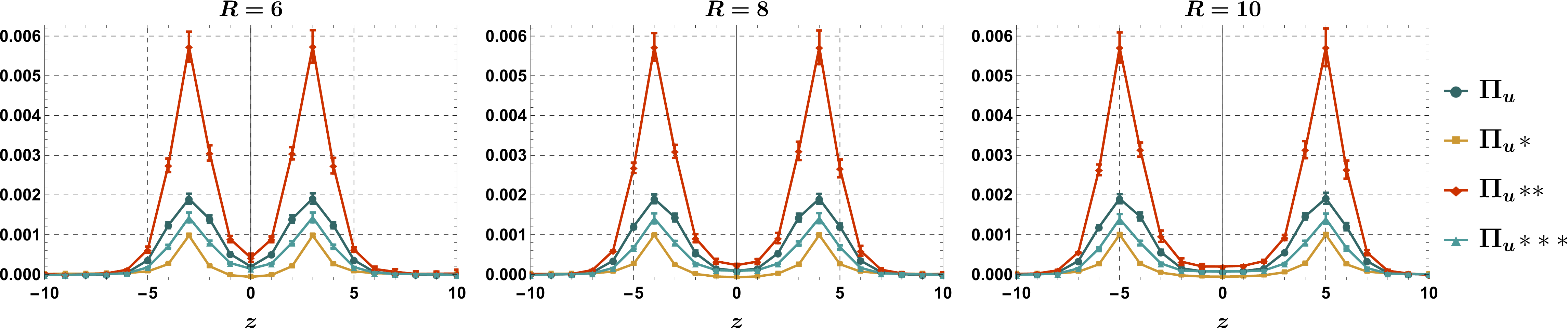}
\par\end{centering}
\caption{Chromoelectric $E^2$ field density in the charges axis. The fields and distances are in lattice spacing units $a$.
\label{fig:field_E_z}}
\end{figure}

\begin{figure}[t!]
\begin{centering}
\includegraphics[width=\LagFigSize]{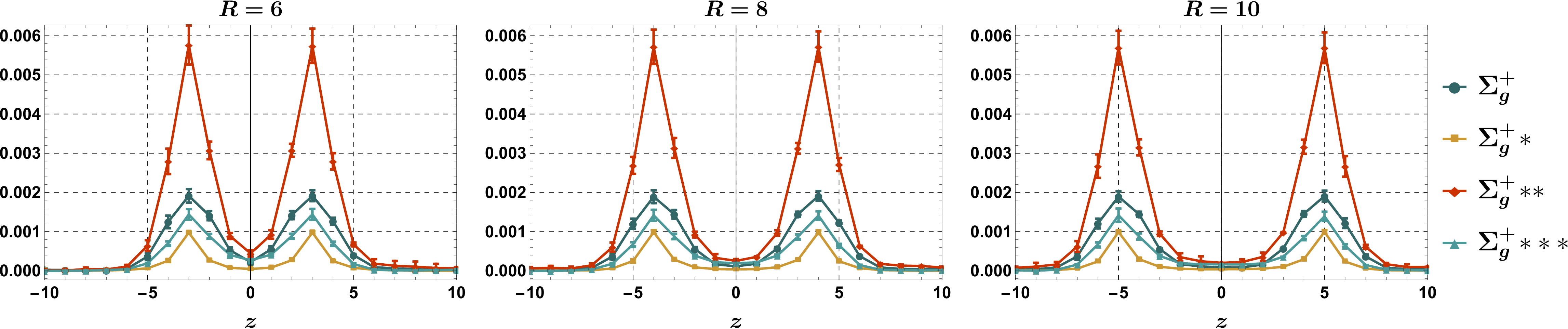}

\includegraphics[width=\LagFigSize]{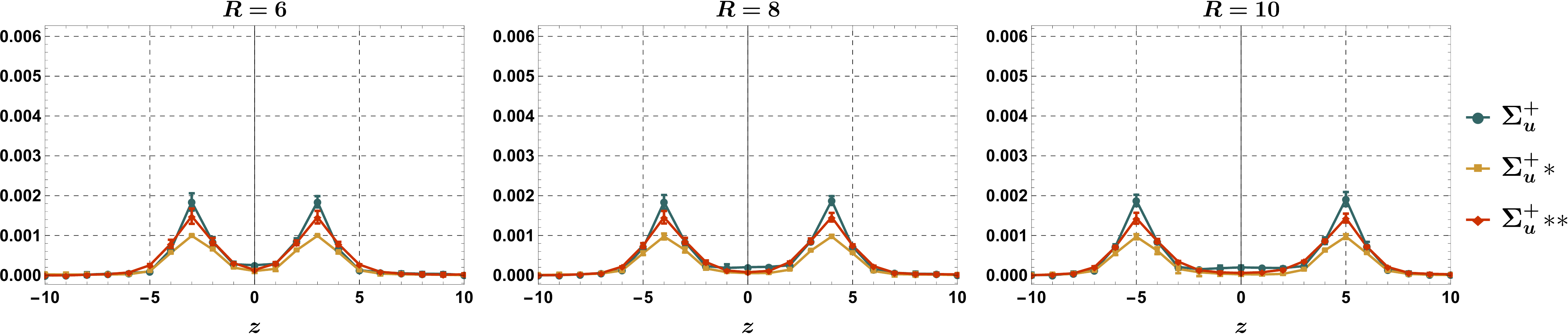}

\includegraphics[width=\LagFigSize]{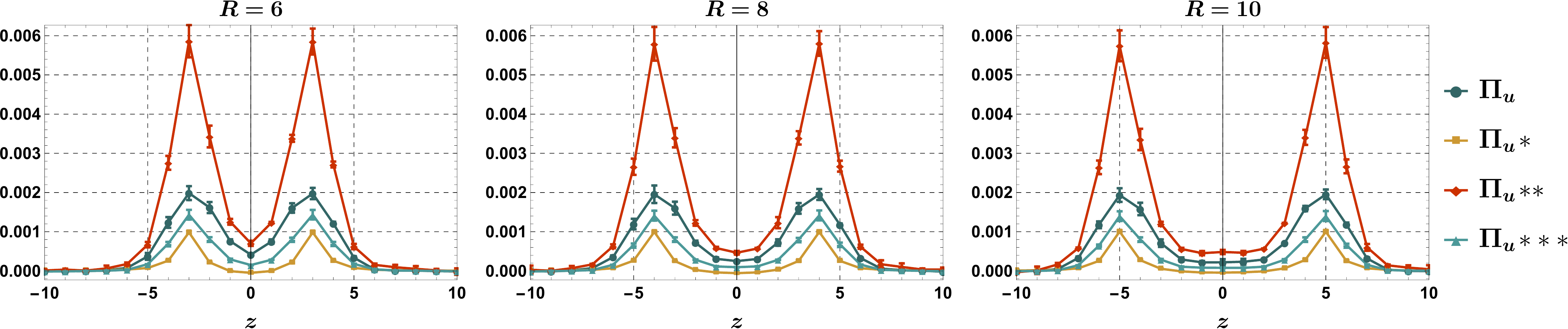}
\par\end{centering}

\caption{Chromomagnetic $B^2$ field density in the charges axis. The fields and distances are in lattice spacing units $a$.
\label{fig:field_B_z}}
\end{figure}

\begin{figure}[t!]
\begin{centering}
\includegraphics[width=\LagFigSize]{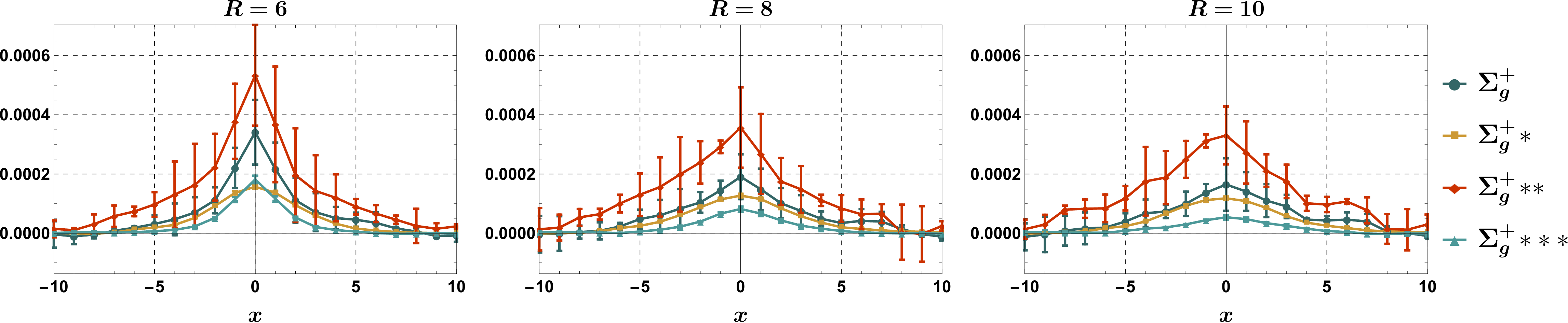}

\includegraphics[width=\LagFigSize]{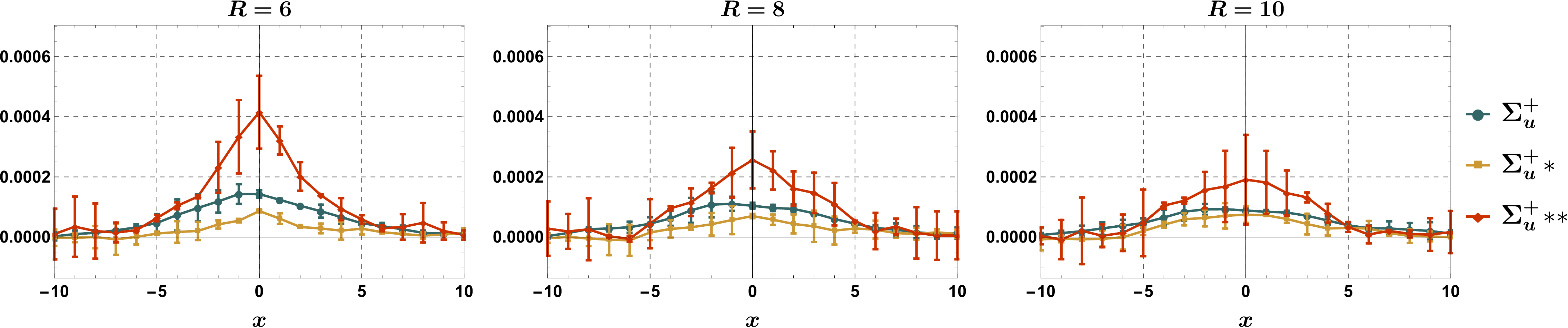}

\includegraphics[width=\LagFigSize]{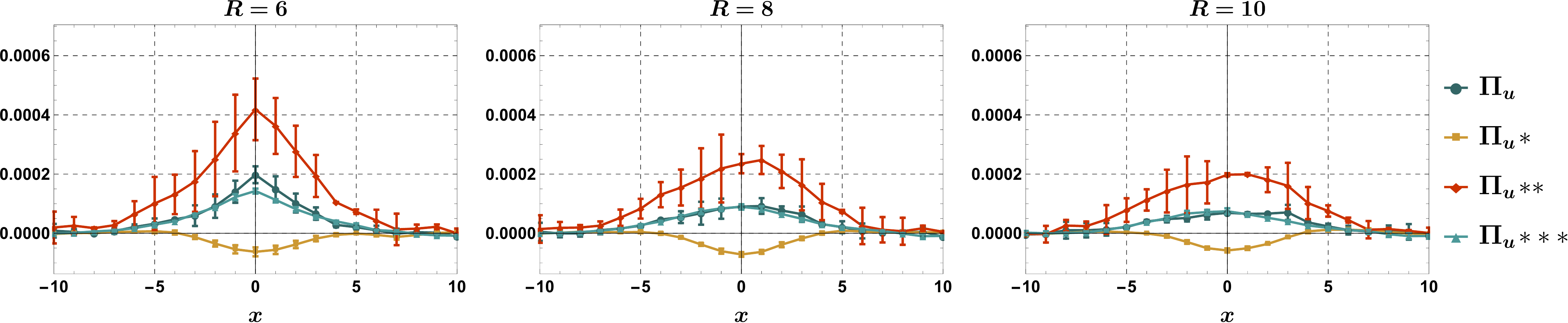}
\par\end{centering}

\caption{Chromoelectric $E^2$ field density in the mediator plane. The fields and distances are in lattice spacing units $a$.
\label{fig:field_E_xy}}
\end{figure}

\begin{figure}[t!]
\begin{centering}
\includegraphics[width=\LagFigSize]{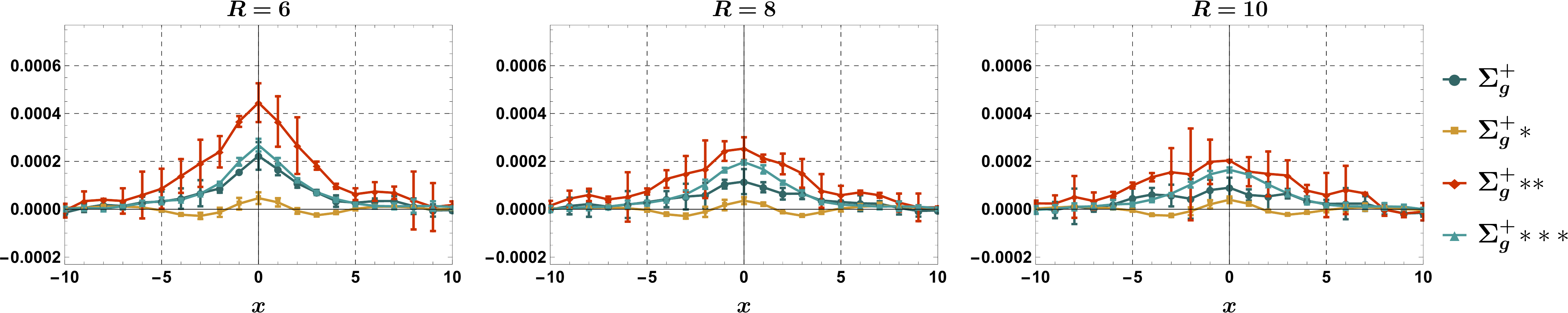}

\includegraphics[width=\LagFigSize]{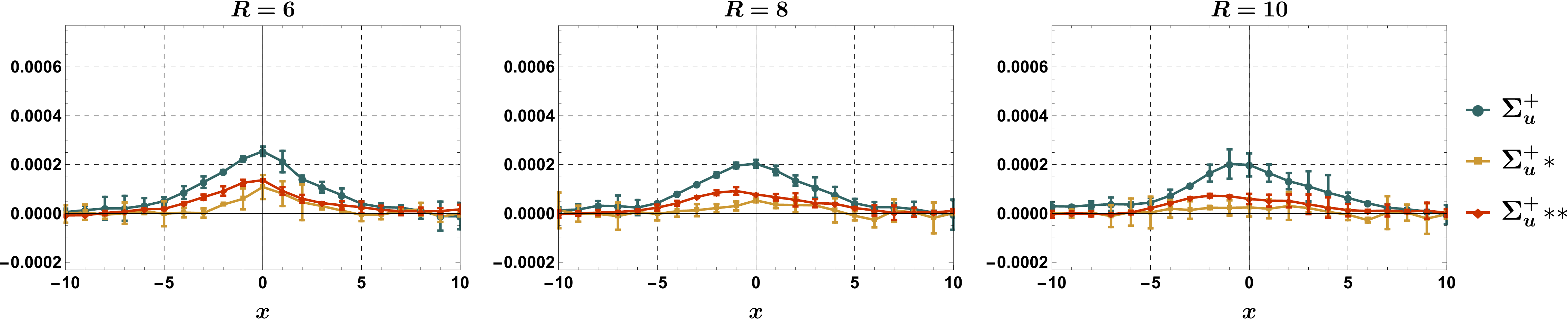}

\includegraphics[width=\LagFigSize]{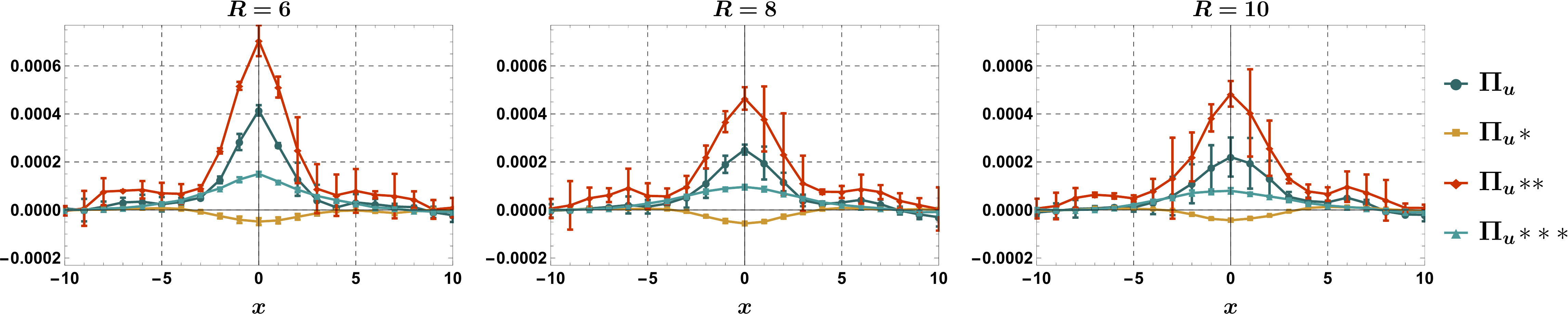}
\par\end{centering}

\caption{Chromomagnetic $B^2$ field density in the mediator plane. The fields and distances are in lattice spacing units $a$.
\label{fig:field_B_xy}}
\end{figure}

Our results for the flux tubes are presented in Figures  \ref{fig:field_E_z}, \ref{fig:field_B_z}, \ref{fig:field_E_xy} and \ref{fig:field_B_xy}. In all these figures we show the flux tubes for the groundstate  $\Sigma^+_g$ and its excitations (s-wave, parity + for the charge conjugation and inversion) and its excitations, the  flux tubes for the $\Sigma_u$ and its excitations (s-wave, parity -   for the charge conjugation and inversion) and the flux tubes for the $\Pi_u$ and its excitations (p-wave,  parity - for the charge conjugation and inversion).

Moreover we show in Figure  \ref{fig:field_E_z} the chromoelectric field density in the charges axis,  in Figure  \ref{fig:field_B_z}  the chromomagnetic field density in the charges axis,  in Figure \ref{fig:field_E_xy}  the chromoelectric field density in the mediator plane, and  in Figure \ref{fig:field_B_xy}  the chromomagnetic field density in the mediator plane.

\section{Conclusions}

We compute the potentials for several excitations of the flux tube.

For the colour field square densities we only select the main excited states in each quantum number. 

We consider radial excitations of the groundstate $\Sigma^+_g$, the first axial parity excitation $\Sigma^+_u$ and the first angular excitation $\Pi_u$.

In our results, Figures  \ref{fig:field_E_z}, \ref{fig:field_B_z}, \ref{fig:field_E_xy} and \ref{fig:field_B_xy} , we compare the chromoelectric and the chromomagnetic field densities, both in the mediator plane and in the charge axis. 

As an outlook, we plan to continue this work, comparing our results with the transverse excitations of the Nambu-Goto string model, but also searching for states who may correspond to excitations beyond the model.

\vspace{0.25cm}
\textbf{Acknowledgments}
\vspace{0.1cm}

Nuno Cardoso and Marco Cardoso are supported by FCT under the contracts SFRH/BPD/109443/2015 and SFRH/BPD/73140/2010 respectively.
We also acknowledge the use of CPU and GPU servers of PtQCD, supported by NVIDIA, CFTP and FCT grant UID/FIS/00777/2013.

\vspace{0.25cm}
\textbf{References}
\vspace{0.1cm}

\bibliographystyle{elsarticle-num}
\bibliography{excited}{}

\end{document}